\documentclass[english,nofootinbib, pra, twocolumn,superscriptaddress]{revtex4-1}
\usepackage[T1]{fontenc}
\usepackage[latin9]{inputenc}
\usepackage{color}
\usepackage{babel}
\usepackage{amsmath}
\usepackage{amssymb}
\usepackage{graphicx}
\usepackage{wasysym}
\usepackage{float}
\usepackage[unicode=true,pdfusetitle,
 bookmarks=true,bookmarksnumbered=false,bookmarksopen=false,
 breaklinks=false,pdfborder={0 0 0},pdfborderstyle={},backref=false,colorlinks=true]
 {hyperref}
\hypersetup{
 citecolor=blue,urlcolor=blue}

\makeatletter
\usepackage{bbm}

\makeatother

\begin{document}
\title{Nonadiabatic evolution and thermodynamics of a time-dependent open quantum
system}
\author{Dan Wang}
\affiliation{Center for Quantum Technology Research and Key Laboratory of Advanced
Optoelectronic Quantum Architecture and Measurements (MOE), School
of Physics, Beijing Institute of Technology, Beijing 100081, China}
\author{Dazhi Xu}
\email{dzxu@bit.edu.cn}

\affiliation{Center for Quantum Technology Research and Key Laboratory of Advanced
Optoelectronic Quantum Architecture and Measurements (MOE), School
of Physics, Beijing Institute of Technology, Beijing 100081, China}
\begin{abstract}
We investigate the dynamic evolution and thermodynamic process of
a driven quantum system immersed in a finite-temperature heat bath.
A Born--Markovian quantum master equation is formally derived for the time-dependent system with discrete energy levels.
This quantum master equation can be applied to situations with a broad range of driving speeds and
bath temperatures and thus be used to study the finite-time quantum thermodynamics even when nonadiabatic transition
and dissipation coexist. The dissipative Landau--Zener model is
analyzed as an example. The population evolution and transition
probability of the model reveal the importance of the competition between driving
and dissipation beyond the adiabatic regime. Moreover, local maximums
of irreversible entropy production occur at intermediate sweep
velocity and finite temperature, which the low-dissipation model cannot describe.
\end{abstract}
\maketitle

\section{Introduction}

The open quantum system with a time-dependent Hamiltonian simultaneously
incorporates coherent driving and unavoidable environment
noise, and can therefore describe a broad range of physical
systems controlled by outside agents. Related problems include
quantum transport \citep{2005.Hanggi}, adiabatic quantum computation
\citep{2002.Preskill,2008.Truncik,2010.Perez,2013.Rose}, optimal
quantum control \citep{2015.DeWeese}, and dissipative many-body systems
\citep{2011.Chen,2015.Clerk,2018.Santoro,2020.Zhai}. In particular,
the recent development of quantum thermodynamics has advanced research
on time-dependent open systems \citep{2009.Mukamel,2011.Talkner,2015.Pekola}.

Although efforts have long been made to develop an appropriate
description of dissipative quantum dynamics with driving \citep{1978.Spohn},
no method is suitable for all the situations. Slowly driven
and periodically driven systems are the two cases studied most, and
formalisms based on the adiabatic master equation \citep{1991.Wudka,2010.Pekola}
and Floquet's theory \citep{1997.Hanggi,2009.Kohn,2015.Esposito,2016.Brandes}
have been widely employed. Moreover, various numerical
methods have been proposed for this issue, such as quasiadiabatic
propagator path integral (QUAPI) \citep{2009.Thorwart}, influence
functional formalism \citep{2017.Kantorovich}, short-iterative Lanczos
method \citep{2019.Filippis}, hierarchical equations of motion \citep{2016.Gong},
quantum trajectories \citep{2018.Lidar}, and variational method \citep{2018.Zhao,2018.Giovannetti}.

However, a general master equation is still needed. This is especially so
when the dissipation and nonadiabatic transition are comparable, a
case that usually occurs for dissipative systems with energy-level avoided
crossings. Assuming that the driving changes slowly
relative to the bath correlation time, a Markovian master equation
has been proposed and applied to the dissipative Landau--Zener (LZ) model
\citep{2014.Nalbach}. However, this assumption is unnecessary, and
the validity of the master equation is not constrained by the adiabatic
condition, as was pointed out in Ref.~\citep{2017.Ogawa} with an elaborate
discussion of the definitions and relations of different dynamic time scales.

In this article, we derive a time-dependent quantum master equation
(TDQME) for systems with discrete energy levels. Adopting Nakajima--Zwanzig projective method \citep{2007.Petruccione}
in the adiabatic frame of reference, this master equation is justified
beyond the adiabatic approximation. The advantage of this approach is
that the nonadiabatic transition and the dissipation are completely
decoupled in the equation; thus, the magnitudes and time scales
of these two crucial factors can be separately considered, which offers great convenience in understanding and even controlling driven dissipative quantum systems. The time evolution and transition probability of the dissipative LZ model are
calculated for demonstration. These results indicate our method can be used beyond the adiabatic condition and verify the conclusions drawn in Ref.~\citep{2017.Ogawa}. 

Furthermore, we can straightforwardly evaluate the average values of thermodynamic quantities, such as heat, work, and entropy, in nonequilibrium processes. The dependence of the irreversible
entropy production on the total evolution time determines the trade-off relation between the efficiency and output power of a heat engine \citep{2010.Broeck,2016.Ryabov,2018.Xu}, which is generally linear in the low-dissipation regime
\citep{2012.Tu,2017.Giovannetti,2020.Dong}. Beyond the low-dissipation regime, we reveal that the irreversible entropy production changes nonmonotonically with the sweep velocity and temperature in the context of the dissipative
LZ model. Therefore, equipped with the TDQME, one can extend the optimal control of the finite-time quantum thermodynamic
processes \citep{2020.Perarnau-Llobet} to the wider parameter regime with profound physical phenomena.

The remainder of the paper is arranged as follows. Section~II formally derives
the TDQME. Section~III presents the example of the dissipative LZ system and the calculation of the dynamics and LZ probability of the system. Section~IV investigates the finite-time thermodynamics of the dissipative LZ problem. Section~V briefly summarizes our results and discusses possible future work. 

\section{TDQME in the adiabatic frame of reference}

In deriving the master equation for a time-independent system, we
usually carry out a secular approximation according to the system
eigenvalues by dropping the fast oscillatory terms. The resulting
dissipation rate $\gamma(E_{ij})$ is then a function of the energy level
difference $E_{ij}$ between the two related eigenstates $\vert i\rangle$
and $\vert j\rangle$ \citep{2007.Petruccione}. This approach is
often generalized to the time-dependent open quantum system by substituting
the time-independent $E_{ij}$ in $\gamma(E_{ij})$ with the instantaneous
eigenvalue difference $E_{ij}(t)$. This naive approach requires that
the system Hamiltonian changes slowly. The meaning of the word ``slowly'' here
is twofolds: 1) the adiabatic theorem \citep{1973.Messiah}
is satisfied such that all the nonadiabatic transitions can be neglected and 2) the system changes slowly relative to the bath correlation time such that the dissipative rate changes following the instantaneous
eigenvalues. Working in the adiabatic frame of reference is a convenient way of deriving a quantum master equation that is justified for an arbitrary driving protocol and a wide range of driving speeds. 

\subsection{Adiabatic frame of reference}

Generally, a time-dependent open system is described by the Hamiltonian
\begin{equation}
H(t)=H_{S}(t)+H_{B}+V,\label{eq:Ht}
\end{equation}
where $H_{S}(t)$ is the system Hamiltonian, $H_{B}$ is the heat bath Hamiltonian, and $V$ describes the system--bath interaction. We assume that the interaction $V$ is weak and all the time-dependent terms are taken
into account by $H_{S}(t)$. The system Hamiltonian can be diagonalized
as
\begin{equation}
H_{S}(t)=\sum_{n}E_{n}(t)\vert n(t)\rangle\langle n(t)\vert,\label{eq:Hst}
\end{equation}
where $\vert n(t)\rangle$ is the instantaneous eigenvector, and $E_{n}(t)$ is 
the corresponding eigenvalue. Under the adiabatic approximation, the
unitary evolution of a time-dependent system with Hamiltonian $H_{S}(t)$ and initial state $\vert n(t_{0})\rangle$ is described as
\begin{equation}
\vert\psi_{n}(t)\rangle=e^{i\eta_{n}(t)+i\xi_{n}(t)}\vert n(t)\rangle,\label{eq:IES}
\end{equation}
where we define the dynamic and adiabatic phase factors 
\begin{eqnarray}
\eta_{n}(t) & = & -\int_{t_{0}}^{t}E_{n}(\tau)d\tau,\label{eq:DF}\\
\xi_{n}(t) & = & i\int_{t_{0}}^{t}\langle n(\tau)\vert\dot{n}(\tau)\rangle d\tau.\label{eq:BF}
\end{eqnarray}
Here, we denote $\vert\dot{n}(\tau)\rangle\equiv\frac{d}{d\tau}\vert n(\tau)\rangle$ and set $\hbar=k_{\mathrm{B}}=1$.

The adiabatic frame of reference is defined by the operator
\begin{equation}
U(t)=\sum_{n}\vert\psi_{n}(t)\rangle\langle n(t_{0})\vert\otimes e^{-iH_{B}t}.\label{eq:Theta}
\end{equation}
The unitarity of $U(t)$ can be easily proved from the orthogonal completeness of $\vert\psi_{n}(t)\rangle$.
$U(t)$ maps the time-dependent state $\vert\psi_{n}(t)\rangle$
to the instantaneous eigenstate $\vert n(t_{0})\rangle$ of
time $t_{0}$, which can be arbitrarily chosen for convenience.
We also incorporate the transformation of the heat bath in $U(t)$;
the transformed total Hamiltonian is then given by
\begin{align}
\tilde{H}(t) & =U^{\dagger}(t)H(t)U(t)-iU^{\dagger}(t)\frac{d}{dt}U(t)\nonumber \\
 & \equiv\tilde{H}_{S}(t)+\tilde{V}(t).\label{eq:Htran}
\end{align}
Here, the transformed system Hamiltonian reads 
\begin{equation}
\tilde{H}_{S}(t)=\sum_{n\neq m}\alpha_{nm}(t)\vert n(t_{0})\rangle\langle m(t_{0})\vert,\label{eq:HsTran}
\end{equation}
where
\begin{equation}
\alpha_{nm}(t)=-ie^{-i\phi_{nm}(t)}\langle n(t)\vert\dot{m}(t)\rangle,\label{eq:a_nm}
\end{equation}
and $\phi_{nm}(t)\equiv\phi_{n}(t)-\phi_{m}(t),$ $\phi_{n}(t)\equiv\eta_{n}(t)+\xi_{n}(t)$.
It is seen from Eq.~(\ref{eq:HsTran}) that $\tilde{H}_{S}(t)$
only includes the nonadiabatic transitions; i.e., the transitions
between the different instantaneous eigenstates. Moreover,
the nonadiabatic transition coefficient $\alpha_{nm}(t)$ is proportional
to $\langle n(t)\vert\dot{m}(t)\rangle$, whose magnitude usually increases
with the driving speed. Therefore, $\tilde{H}_{S}(t)$ can be neglected
when the adiabatic theorem is satisfied. Otherwise, we should consider
its contribution to the dissipative dynamics of the driven open quantum
system. 

The transformed system--bath interaction is written as
\begin{equation}
\tilde{V}(t)=U^{\dagger}(t)VU(t),\label{eq:Vt}
\end{equation}
which is now time dependent. We should keep all the rotating wave and counter-rotating wave terms in the
system--bath coupling $V$ to ensure the secular approximation
is carried out correctly in the adiabatic frame of reference.

\subsection{Nakajima--Zwanzig projective approach}

We follow the standard Nakajima--Zwanzig projective operator method
\citep{2007.Petruccione} to derive the TDQME, such that the nonadiabatic
transition and dissipation can be considered in the same framework.
In the adiabatic frame of reference defined above, the quantum Liouville
equation reads 
\begin{equation}
\frac{d}{dt}\tilde{\rho}(t)=-i[\tilde{H}(t),\tilde{\rho}(t)]\equiv\mathcal{L}(t)[\tilde{\rho}(t)].\label{eq:Liouville}
\end{equation}
Here, $\tilde{\rho}(t)$ is the density operator of the total system
in the adiabatic frame and relates to the operator $\rho(t)$ in the
original frame according to $\tilde{\rho}(t)=U^{\dagger}(t)\rho(t)U(t)$. The
Liouville superoperator $\mathcal{L}(t)$ can be decomposed into two
parts $\mathcal{L}=\mathcal{L}_{\text{0}}+\mathcal{L}_{I}$, where
\begin{eqnarray}
\mathcal{L}_{\text{0}}[\tilde{\rho}(t)] & = & -i[\tilde{H}_{S}(t),\tilde{\rho}(t)],\label{eq:Lnon}\\
\mathcal{L}_{I}[\tilde{\rho}(t)] & = & -i[\tilde{V}(t),\tilde{\rho}(t)].
\end{eqnarray}
Then, $\mathcal{L}_{\text{0}}$ describes the nonadiabatic evolution while $\mathcal{L}_{I}$ describes the system--bath interaction that should be treated perturbatively.

The projective operator $\mathcal{P}$ is defined as 
\begin{equation}
\mathcal{P}\tilde{\rho}(t)=\mbox{Tr}_{B}[\tilde{\rho}(t)]\otimes\rho_{B}\equiv\tilde{\rho}_{S}(t)\otimes\rho_{B}.\label{eq:P}
\end{equation}
We see that $\mathcal{P}$ maps $\tilde{\rho}(t)$ to a direct product of
the reduced-density operator of the system $\tilde{\rho}_{S}(t)$ and that of the 
heat bath $\rho_{B}$. As the system--bath coupling is weak, it is
reasonable to assume that the heat bath is always in thermal equilibrium; i.e., $\rho_{B}=\exp(-\beta H_{B})/\text{Tr}[\exp(-\beta H_{B})]$. The orthogonal projective operator $\mathcal{Q}$ is defined by 
\begin{equation}
\mathcal{Q}\tilde{\rho}(t)=\tilde{\rho}(t)-\mathcal{P}\tilde{\rho}(t).\label{eq:Q}
\end{equation}
$\mathcal{P}$ and $\mathcal{Q}$ satisfy the relations $\mathcal{P}^{2}=\mathcal{P}$,
$\mathcal{Q}^{2}=\mathcal{Q}$ and $\mathcal{P}\mathcal{Q}=\mathcal{Q}\mathcal{P}=0$.
We also assume that the expectation value of odd orders of $\tilde{V}(t)$
is zero, which can be written with superoperators as $
\mathcal{P}\mathcal{L}_{I}(t_{1})\mathcal{L}_{I}(t_{2})\cdots\mathcal{L}_{I}(t_{2n+1})\mathcal{P}=0.$

The quantum Liouville equation can be decomposed into two coupled
equations by applying $\mathcal{P}$ and $\mathcal{Q}$ to Eq.~(\ref{eq:Liouville}):
\begin{eqnarray}
\frac{d}{dt}\mathcal{P}\tilde{\rho}(t) & = & \mathcal{P}\mathcal{L}(t)\mathcal{P}\tilde{\rho}(t)+\mathcal{P}\mathcal{L}(t)\mathcal{Q}\tilde{\rho}(t),\label{eq:Pequ}\\
\frac{d}{dt}\mathcal{Q}\tilde{\rho}(t) & = & \mathcal{Q}\mathcal{L}(t)\mathcal{P}\tilde{\rho}(t)+\mathcal{Q}\mathcal{L}(t)\mathcal{Q}\tilde{\rho}(t).\label{eq:Qequ}
\end{eqnarray}
Formally integrating Eq.~(\ref{eq:Qequ}), we obtain
\begin{equation}
\mathcal{Q}\tilde{\rho}(t)=\mathcal{G}(t,t_{0})\mathcal{Q}\tilde{\rho}(t_{0})+\int_{t_{0}}^{t}ds\mathcal{G}(t,s)\mathcal{Q}\mathcal{L}(s)\mathcal{P}\tilde{\rho}(s),\label{eq:Qint}
\end{equation}
where Green's function $\mathcal{G}\left(t,s\right)$ is defined as
\begin{equation}t
\mathcal{G}(t,s)=\mbox{T}_{\leftarrow}\exp[\int_{s}^{t}d\tau\mathcal{Q}\mathcal{L}(\tau)],
\end{equation}
and $\mbox{T}_{\leftarrow}$ is the chronological time operator indicating that
the time argument increases from right to left. Substituting Eq.~(\ref{eq:Qint})
into Eq.~(\ref{eq:Pequ}) gives
\begin{equation}
\frac{d}{dt}\mathcal{P}\tilde{\rho}(t)=\mathcal{P}\mathcal{L}(t)\mathcal{P}\tilde{\rho}(t)+\int_{t_{0}}^{t}ds\mathcal{K}(t,s)\tilde{\rho}(s),\label{eq:N-Zequ}
\end{equation}
where we define the convolution kernel operator as
\begin{equation}
\mathcal{K}(t,s)=\mathcal{P}\mathcal{L}(t)\mathcal{G}(t,s)\mathcal{Q}\mathcal{L}(s)\mathcal{P}.\label{eq:kernel}
\end{equation}
In deriving Eq.~(\ref{eq:N-Zequ}), we also adopt the Born approximation
by assuming the initial state is a direct product state $\tilde{\rho}(t_{0})=\tilde{\rho}_{S}(t_{0})\otimes\rho_{B}$,
such that $\mathcal{Q}\tilde{\rho}(t_{0})=0$. Using $\mathcal{P}\mathcal{L}_{I}(t)\mathcal{P}=0$
and $\mathcal{L}_{\text{0}}(t)\mathcal{Q}=\mathcal{Q}\mathcal{L}_{\text{0}}(t)$,
we only keep the master
equation to the second order of $\mathcal{L}_{I}$ and approximately write
\begin{equation}
\mathcal{K}(t,s)\approx\mathcal{P}\mathcal{L}_{I}(t)\mbox{T}_{\leftarrow}e^{\int_{s}^{t}d\tau\mathcal{L}_{\text{0}}(\tau)}\mathcal{L}_{I}(s)\mathcal{P}.\label{eq:K}
\end{equation}

Furthermore, we adopt the Markovian approximation by substituting the integral
variable $s\rightarrow t-s$ in Eq.~(\ref{eq:N-Zequ}) and Eq.~(\ref{eq:K}),
and assume $\tilde{\rho}(t-s)\approx\tilde{\rho}(t)$ considering
that the system density operator changes slightly within the time
scale of the bath correlation time $\tau_{\text{B}}$. We thus have
\begin{align}
 & \int_{t_{0}}^{t}ds\mathcal{K}(t,s)\tilde{\rho}(s)\nonumber \\
\approx & \int_{0}^{t-t_{0}}ds\mathcal{P}\mathcal{L}_{I}(t)\mbox{T}_{\rightarrow}e^{\int_{0}^{s}d\tau\mathcal{L}_{\text{0}}(t-\tau)}\mathcal{L}_{I}(t-s)\mathcal{P}\tilde{\rho}(t).\label{eq:NZeq}
\end{align}
The exponential superoperator on the right side of the above equation
can be written explicitly as
\begin{equation}
\mbox{T}_{\rightarrow}e^{\int_{0}^{s}d\tau\mathcal{L}_{\text{0}}(t-\tau)}\hat{O}=\tilde{R}(s)\hat{O}\tilde{R}^{\dagger}(s),
\end{equation}
where $\hat{O}$ is an arbitrary operator and 
\begin{equation}
\tilde{R}(s)=\mbox{T}_{\rightarrow}\exp[-i\int_{0}^{s}d\tau\tilde{H}_{S}(t-\tau)].
\end{equation}
As seen from the dissipative LZ model below, Eq.~(\ref{eq:NZeq}) is proportional to
the bath correlation function, which quickly decays within the short
time scale $\tau_{\text{B}}$. Therefore,
$\tilde{R}(s)$ contributes to Eq.~(\ref{eq:NZeq}) mainly during
$0\lesssim s\lesssim\tau_{\text{B}}$, such that we can safely approximate
$\tilde{R}(s)\approx1$, which is equivalent to assuming that
\begin{equation}
\vert\langle n(t)\vert\dot{m}(t)\rangle\vert\ll\tau_{\text{B}}^{-1}.\label{eq:condition}
\end{equation}
This condition is generally easy to be satisfied, even if the adiabatic condition
\citep{1973.Messiah} is no longer justified. 

Combining Eqs.~(\ref{eq:N-Zequ},\ref{eq:NZeq},\ref{eq:condition})
and tracing over the bath degrees of freedom, we finally obtain the
general form of the TDQME: 
\begin{equation}
\frac{d}{dt}\tilde{\rho}_{S}(t)=\mathcal{L}_{\text{0}}[\tilde{\rho}_{S}(t)]+\mathcal{L}_{\text{B}}[\tilde{\rho}_{S}(t)],\label{eq:TDQME}
\end{equation}
where $\mathcal{L}_{\text{0}}$ is defined by Eq.~(\ref{eq:Lnon})
and $\mathcal{L}_{\text{B}}$ is expressed as
\begin{equation}
\mathcal{L}_{\text{B}}[\tilde{\rho}_{S}(t)]=-\int_{0}^{t-t_{0}}ds\textrm{Tr}_{\text{B}}[\tilde{V}(t),[\tilde{V}(t-s),\tilde{\rho}_{S}(t)\otimes\rho_{B}]].\label{eq:LB}
\end{equation}
The reduced density operator of the system can be obtained by solving
Eq.~(\ref{eq:TDQME}) and then transforming back into the original frame
using $\rho_{S}(t)=\text{Tr}_{B}[U(t)\tilde{\rho}_{S}(t)U^{\dagger}(t)]$.
The transformation $U(t)$ takes the responsibility of the adiabatic evolution.
If there is no nonadiabatic transition or dissipation, $U(t)$ adiabatically rotates the system following the instantaneous eigenstates.

It is noted that, as indicated in Ref.~\citep{2017.Ogawa},
the TDQME in the weak system--bath coupling case is actually not constrained
by the condition (\ref{eq:condition}) and can be applied to
a situation with larger nonadiabatic transitions. Putting it briefly, the violation of (\ref{eq:condition})
implies that the nonadiabatic transition is strong, such that the
$\mathcal{L}_{\text{B}}$ is negligible regardless of how it is derived. We will illustrate this point in the following
section taking the example of the dissipative LZ problem.

\section{Dissipative LZ system}

\subsection{TDQME for a linearly driven two-level system}

\begin{figure}
\includegraphics[width=7cm]{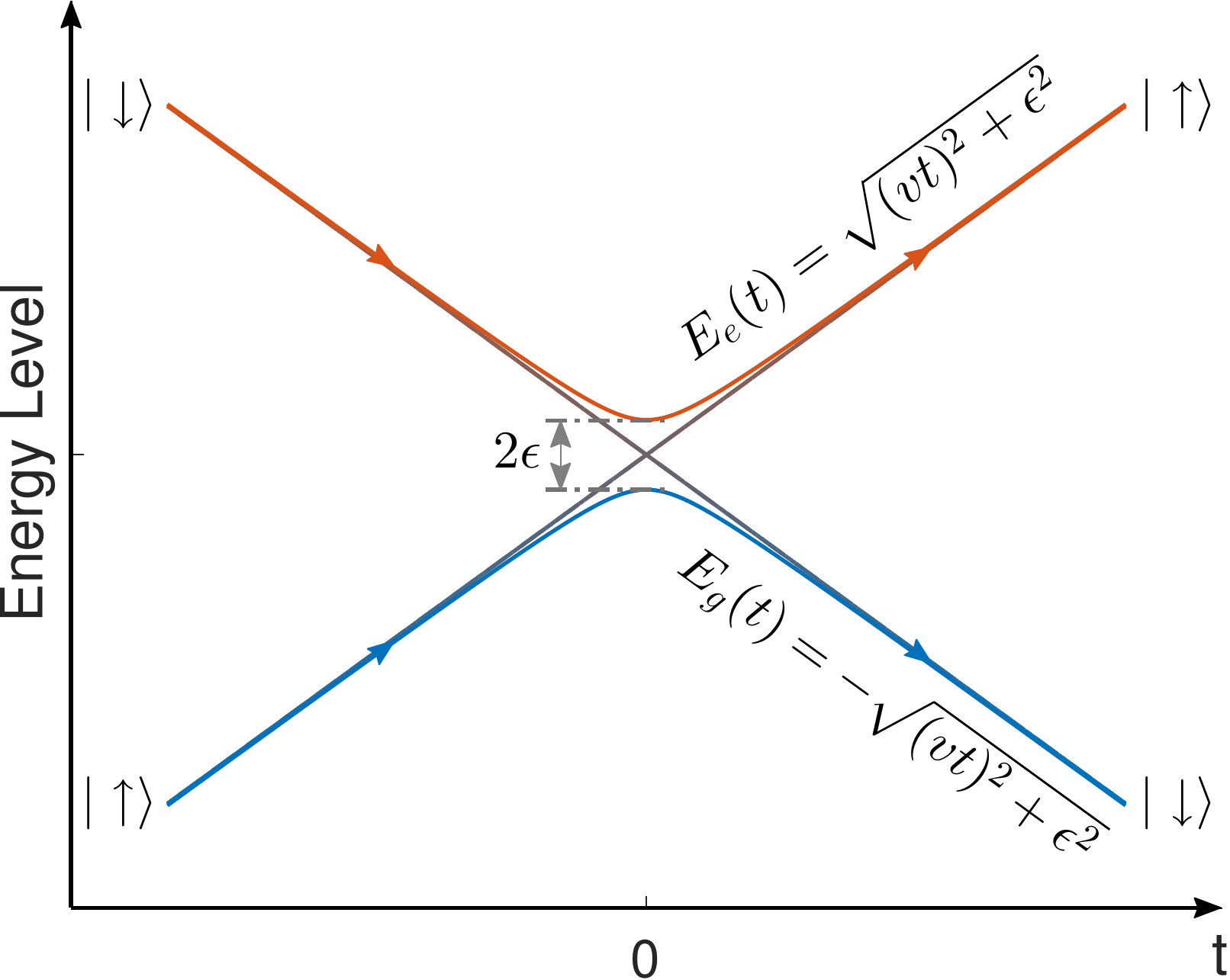}\caption{Schematic illustration of the LZ model. The minimal energy gap at
$t=0$ is $2\epsilon$. The time-dependent eigenvalues $E_{e,g}(t)$
are plotted with solid lines. In the adiabatic regime, the basis of
the system transform from $\vert\uparrow\rangle(\vert\downarrow\rangle)$
at $t=-\infty$ to $\vert\downarrow\rangle(\vert\uparrow\rangle)$
at $t=\infty$ following the instantaneous eigenstates.}
\label{fig:LZ}
\end{figure}

The linearly driven two-level system interacting with a bosonic bath
is an ideal model with which to demonstrate how to apply the TDQME,
as it is both simple enough yet contains almost all the essential
ingredients of dissipative open systems. This model is also a good
playground in which to study finite-time quantum thermodynamics.
If we focus on the physics with avoided crossing point, this model becomes the well-known LZ model with the Hamiltonian
\begin{equation}
H_{S}(t)=vt\sigma_{z}+\epsilon\sigma_{x},\label{eq:Hs}
\end{equation}
where $v$ is the sweep velocity and $\epsilon$ is the tunneling
amplitude. The Pauli operators $\sigma_{x,z}$ are defined in the
diabatic basis representation $\vert\uparrow\rangle$ and $\vert\downarrow\rangle$
as $\sigma_{x}=\vert\uparrow\rangle\langle\downarrow\vert+\vert\uparrow\rangle\langle\downarrow\vert$
and $\sigma_{z}=\vert\uparrow\rangle\langle\uparrow\vert-\vert\downarrow\rangle\langle\downarrow\vert$.
The instantaneous eigenstates are 
\begin{eqnarray}
\vert e(t)\rangle & = & \cos\frac{\theta_{t}}{2}\vert\uparrow\rangle+\sin\frac{\theta_{t}}{2}\vert\downarrow\rangle,\label{eq:ES+}\\
\vert g(t)\rangle & = & -\sin\frac{\theta_{t}}{2}\vert\uparrow\rangle+\cos\frac{\theta_{t}}{2}\vert\downarrow\rangle,\label{eq:ES-}
\end{eqnarray}
where $\tan\theta_{t}=\epsilon/(vt)$, and the corresponding eigenvalues are
\begin{equation}
E_{e}(t)=-E_{g}(t)=\sqrt{v^{2}t^{2}+\epsilon^{2}}.\label{eq:EV}
\end{equation}
The instantaneous energy levels are illustrated in Fig.~(\ref{fig:LZ}).
It is easy to show that Berry's phases $\xi_{e,g}(t)$ are all zero:
\begin{equation}
\langle e(t)\vert\dot{e}(t)\rangle=\langle g(t)\vert\dot{g}(t)\rangle=0.\label{eq:BF_LZ}
\end{equation}
The effect of a finite Berry's phase on the dissipative dynamics is
another interesting problem \citep{2005.Gefen,2007.Palma} but is
not discussed in the current paper. The strength of the nonadiabatic transition
(or the LZ transition) is expressed as
\begin{equation}
\langle e(t)\vert\dot{g}(t)\rangle=-\langle g(t)\vert\dot{e}(t)\rangle=\frac{\epsilon v/2}{v^{2}t^{2}+\epsilon^{2}}.\label{eq:DiaTransi_LZ}
\end{equation}
Equation~(\ref{eq:DiaTransi_LZ}) is a Lorentzian function
of $t$ with maximum $v/(2\epsilon)$ and width $2\tau_{\text{LZ}}$,
where $\tau_{\text{LZ}}=\epsilon/v$ is the time scale on which
the LZ transition dominates the dynamics. If the sweep velocity is
high, the nonadiabatic transition appears like a pulse with
large amplitude and width $2\tau_{\text{LZ}}$.

The bath Hamiltonian $H_{B}$ and the system--bath coupling $V$ are expressed as
\begin{equation}
H_{B}=\sum_{k}\omega_{k}a_{k}^{\dagger}a_{k},\label{eq:HB}
\end{equation}
and
\begin{equation}
V=\sigma_{x}\otimes\sum_{k}g_{k}(a_{k}^{\dagger}+a_{k}),\label{eq:VLZ}
\end{equation}
where $a_{k}$ is the bosonic annihilation operator of the bath harmonic
mode with frequency $\omega_{k}$ while $g_{k}$ is the coupling
strength. For simplicity, we consider only the case that the bath is transversely
coupled with the two-level system via operator $\sigma_{x}$ in the main text. The longitudinal $\sigma_{z}$-coupling case is briefly discussed in Appendix~\ref{sec:A}. The linear combination of
these two kinds of coupling was discussed in Ref.~\citep{2015.Thorwart}. 

Applying the unitary transformation defined in (\ref{eq:Theta}),
we obtain
\begin{equation}
\tilde{H}_{S}(t)=\alpha_{eg}(t)\hat{\sigma}_{+}+\text{h.c.},\label{eq:HSLZ}
\end{equation}
and 
\begin{equation}
\tilde{V}(t)=\tilde{\sigma}_{x}(t)\otimes B(t),\label{eq:Vad}
\end{equation}
where $B(t)=\sum_{k}g_{k}(a_{k}^{\dagger}e^{i\omega_{k}t}+a_{k}e^{-i\omega_{k}t})$
and 
\begin{equation}
\tilde{\sigma}_{x}(t)=\sin\theta_{t}\hat{\sigma}_{z}+\cos\theta_{t}[e^{-i\phi_{eg}(t)}\hat{\sigma}_{+}+\text{h.c.}].\label{eq:Sxad}
\end{equation}
Here, the Pauli operators $\hat{\sigma}_{x,y,z,\pm}$ are defined in
the presentation with the basis $\{\vert e(t_{0})\rangle,\vert g(t_{0})\rangle\}$; e.g., $\hat{\sigma}_{+}=\vert e(t_{0})\rangle\langle g(t_{0})\vert$.

After substituting Eqs.~(\ref{eq:Vad},\ref{eq:Sxad}) into Eq.~(\ref{eq:LB}),
we can further use the secular approximation to neglect the fast oscillating terms, such as those containing the factor
$e^{\pm i\phi_{eg}(t)}$ or $e^{\pm i\phi_{eg}(t-s)}$. The dissipation
superoperator can then be expressed as
\begin{eqnarray}
\mathcal{L}_{\text{B}}[\tilde{\rho}_{S}(t)] & = & -i\left[S_{+}(t)\hat{\sigma}_{+}\hat{\sigma}_{-}+S_{-}(t)\hat{\sigma}_{-}\hat{\sigma}_{+},\tilde{\rho}_{S}(t)\right]\nonumber \\
 &  & -\Gamma_{+}(t)\left(\left\{ \hat{\sigma}_{+}\hat{\sigma}_{-},\tilde{\rho}_{S}(t)\right\} -2\hat{\sigma}_{-}\tilde{\rho}_{S}(t)\hat{\sigma}_{+}\right)\nonumber \\
 &  & -\Gamma_{-}(t)\left(\left\{ \hat{\sigma}_{-}\hat{\sigma}_{+},\tilde{\rho}_{S}(t)\right\} -2\hat{\sigma}_{+}\tilde{\rho}_{S}(t)\hat{\sigma}_{-}\right)\nonumber \\
 &  & -2\Gamma_{z}(t)\left[\tilde{\rho}_{S}(t)-\hat{\sigma}_{z}\tilde{\rho}_{S}(t)\hat{\sigma}_{z}\right],\label{eq:LBLZ}
\end{eqnarray}
where the energy-level shift $S_{\pm}(t)$, relaxation rate $\Gamma_{\pm}(t)$,
and dephasing rate $\Gamma_{z}(t)$ are given by
\begin{eqnarray}
S_{\pm}(t) & = & \cos\theta_{t}\int_{0}^{t-t_{0}}ds\cos\theta_{t-s}\text{Im}[C(s)e^{\pm i\Delta(t,s)}],\label{eq:Spm}\\
\Gamma_{\pm}(t) & = & \cos\theta_{t}\int_{0}^{t-t_{0}}ds\cos\theta_{t-s}\text{Re}[C(s)e^{\pm i\Delta(t,s)}],\label{eq:G+-}\\
\Gamma_{z}(t) & = & \sin\theta_{t}\int_{0}^{t-t_{0}}ds\sin\theta_{t-s}\text{Re}[C(s)].\label{eq:Gz}
\end{eqnarray}
Here $\Delta(t,s)=2\int_{0}^{s}d\tau\sqrt{v^{2}(t-\tau)^{2}+\epsilon^{2}}$
and $C(s)$ is the bath correlation function
\begin{align}
C(s) & \equiv\textrm{Tr}_{\text{B}}[B(t)B(t-s)\rho_{B}]\nonumber \\
 & =\frac{1}{\pi}\int_{0}^{\infty}d\omega J(\omega)\left[\coth\left(\frac{\beta\omega}{2}\right)\cos(\omega s)-i\sin(\omega s)\right],\label{eq:cs1}
\end{align}
where the system-bath coupling spectral density is defined as $J(\omega)=\pi\sum g_{k}^{2}\delta(\omega-\omega_{k})$.
In this work, we assume that $J(\omega)$ is an Ohmic spectrum with an
exponential cutoff $J(\omega)=\pi\gamma\omega e^{-\omega/\omega_{c}}$, where
$\gamma$ is a constant characterizing the system--bath coupling strength,
and $\omega_{c}$ is the cutoff frequency. With such settings,
the bath correlation function can be calculated as 
\begin{equation}
C(s)=\gamma\omega_{c}^{2}\left[\frac{2\text{Re}[\psi^{(1)}(\frac{1}{\beta\omega_{c}}+\frac{is}{\beta})]}{(\beta\omega_{c})^{2}}+\frac{1}{(s\omega_{c}+i)^{2}}\right]\label{eq:cs2}
\end{equation}
with $\psi^{(1)}(z)$ the trigamma function. We can also choose $J(\omega)$
to be an Ohmic spectrum with a Lorentz--Drude cutoff and obtain the analytical
form of $C(s)$ with the help of the Matsubara expansion, which does not
introduce any qualitative difference from the current work. However, the convergence
of the Matsubara expansion is much slower at low temperature.

\subsection{Time evolution}

The two most important and competitive ingredients of the dissipative
LZ system, namely the LZ transition and dissipation, are now separately described
in Eq.~(\ref{eq:TDQME}) by the unitary evolution $\mathcal{L}_{\text{0}}[\tilde{\rho}_{S}(t)]$
and non-unitary evolution $\mathcal{L}_{\text{B}}[\tilde{\rho}_{S}(t)]$,
respectively. The unraveling of the dynamic processes allows us to
clarify how the LZ transition and dissipation contribute on different
time scales. 

\begin{figure}
\includegraphics[width=8cm]{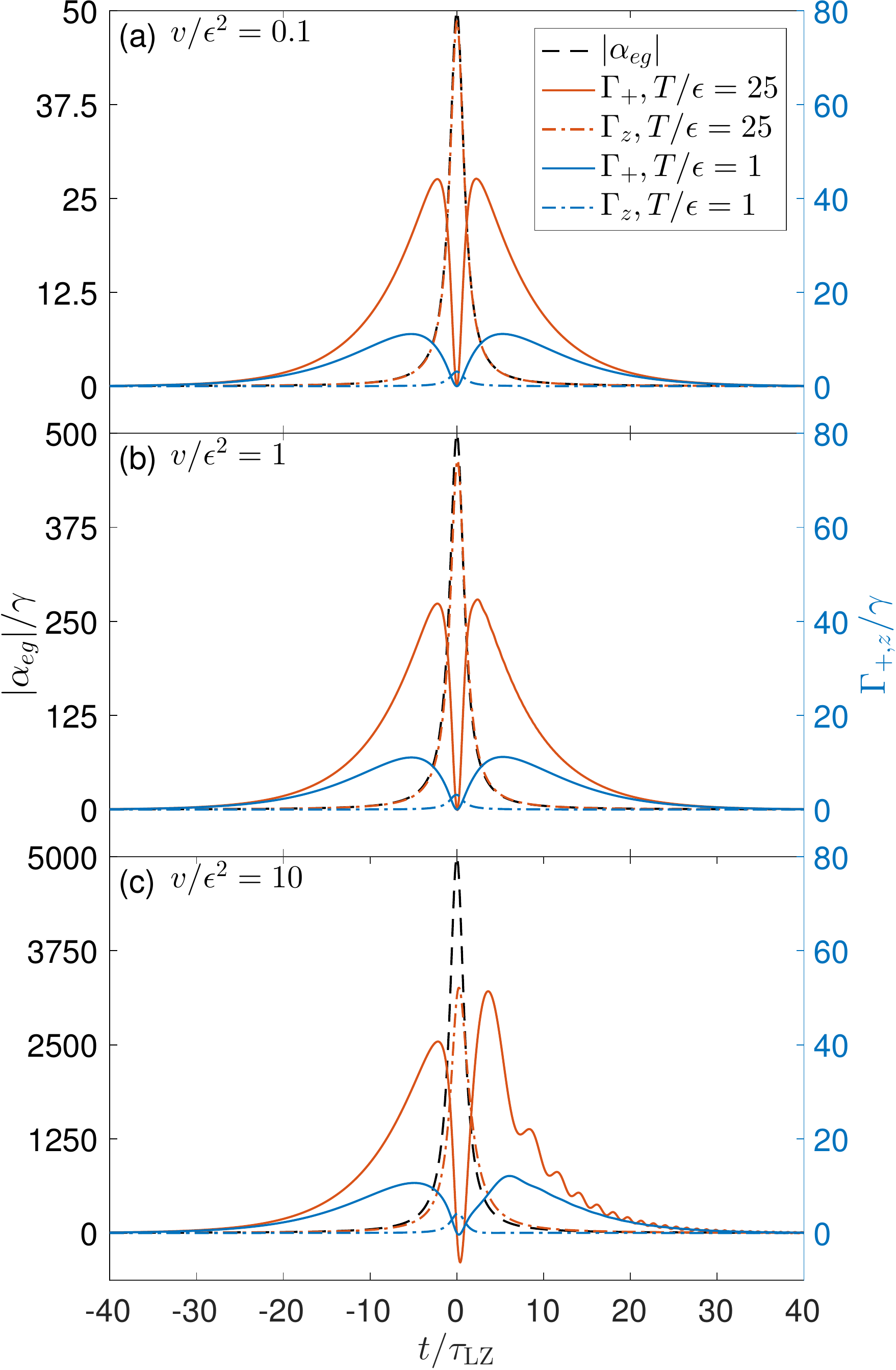}\caption{Strength of the nonadiabatic transition $\vert\alpha_{eg}(t)\vert$,
relaxation rate $\Gamma_{+}(t)$, and dephasing rate $\Gamma_{z}(t)$
as functions of the evolution time $t$. Different sweep velocities are
chosen in the (a) slow ($v/\epsilon^{2}=0.1$), (b) intermediate ($v/\epsilon^{2}=1$),
and (c) fast ($v/\epsilon^{2}=10$) regimes. The nonadiabatic transition
$\vert\alpha_{eg}(t)\vert$ (black dashed line) is a Lorentzian peak
with width $2\tau_{\text{LZ}}$ and maximum $v/(2\epsilon)$. The
relaxation rate $\Gamma_{+}(t)$ has a dip around $t=0$, and its magnitude decreases with temperature
($T/\epsilon=25$ for the orange solid line and $T/\epsilon=1$ for the blue
solid line). The dephasing rate $\Gamma_{z}(t)$ (dot-dashed lines) appears
on the same time scale as $\vert\alpha_{eg}(t)\vert$, but its magnitude
is much smaller. Here, we set $\gamma=0.001\epsilon$ and $\omega_{c}=10\epsilon$. }
\label{fig:decay}
\end{figure}

We plot $\vert\alpha_{eg}(t)\vert$, $\Gamma_{+}(t)$ and $\Gamma_{z}(t)$
as functions of the rescaled time $t/\tau_{\text{LZ}}$ in Fig.~\ref{fig:decay}
for different sweep velocities $v/\epsilon^{2}=0.1,1$ and $10$,
at both high temperature $T=25\epsilon$ and low temperature
$T=\epsilon$. Both $\Gamma_{+}(t)$ and $\Gamma_{z}(t)$ decrease
with temperature, while $\vert\alpha_{eg}(t)\vert$ is independent of temperature.
Consistent with our discussion in Sec.~IIIA, the Lorentzian line shape of $\vert\alpha_{eg}(t)\vert$ lies at the avoided crossing
point $t=0$, where there is a sharp decline of $\Gamma_{+}(t)$. The
height of $\vert\alpha_{eg}(t)\vert$ linearly increases with the sweep
velocity. Entering the nonadiabatic regime $v/\epsilon^{2}>1$,
the LZ transition overwhelms the dissipation. The relaxation rate
$\Gamma_{+}(t)$ has two maximums located around $t\approx\pm2\tau_{\text{LZ}}$,
beyond which $\Gamma_{+}(t)$ exponentially decreases to zero on a
time scale of about $2\omega_{c}/v$. The curves of $\Gamma_{-}(t)$
are similar to but lower than those of $\Gamma_{+}(t)$ and are not presented here. 

\begin{figure*}
\includegraphics[width=13cm]{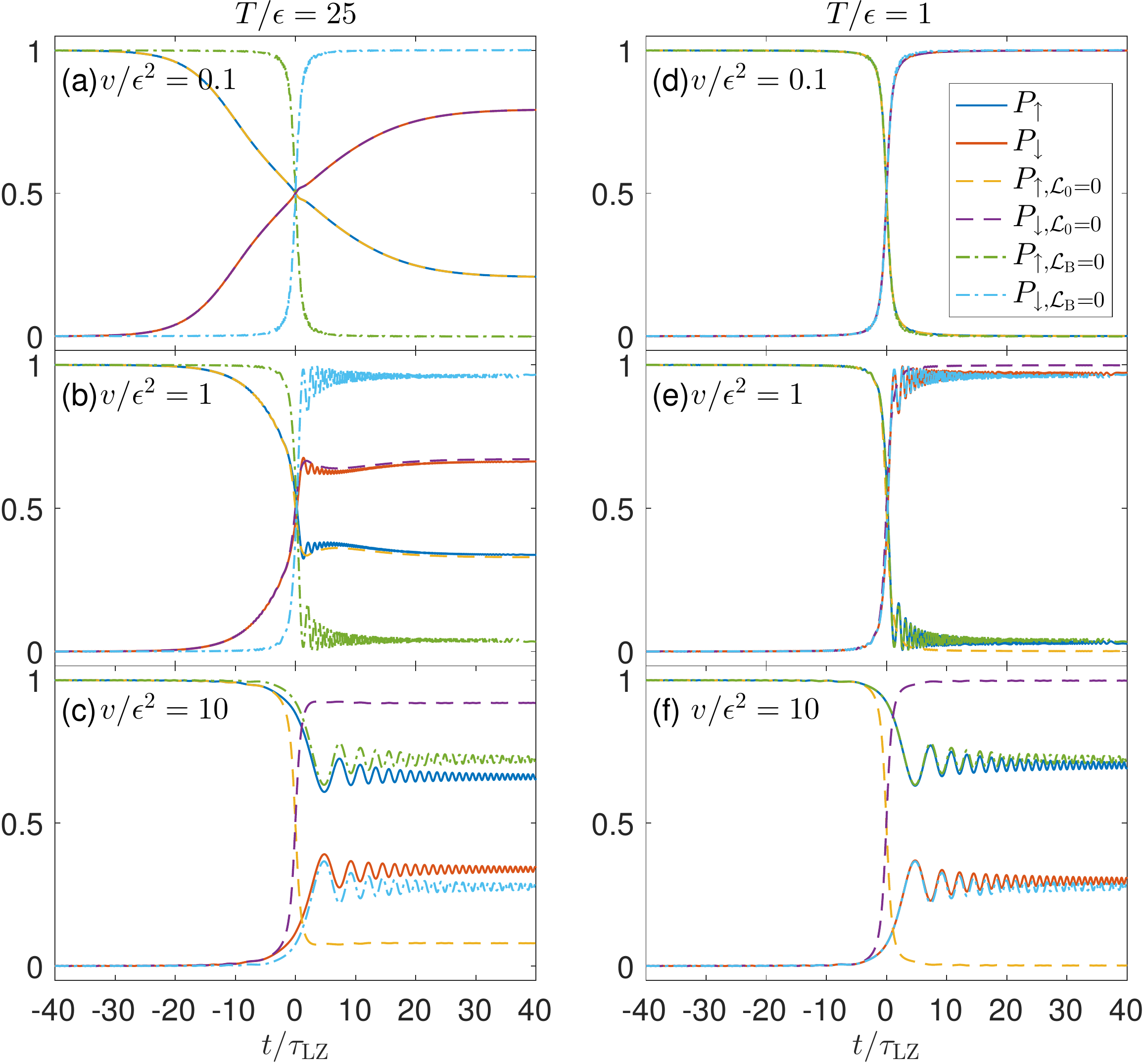}
\caption{Time evolution of diabatic-state populations of the dissipative LZ
model. The initial state is chosen as $\vert\uparrow\rangle$. The
evolution starts at $t_{0}=-40\tau_{\text{LZ}}$ and ends at $t_{f}=40\tau_{\text{LZ}}$.
From top to bottom, the two subplots in each row correspond to $v/\epsilon^{2}=0.1$,
$1$, and $10$; the left (right) panel is the high (low)
temperature case with $T/\epsilon=25$ ($T/\epsilon=1$). The populations
$P_{\uparrow,\downarrow}(t)$ calculated using the TDQME are plotted
with solid lines. For comparison, we also plot the populations evolutions
in the cases without the nonadiabatic transition ($\mathcal{L}_{\text{0}}=0$)
and without the dissipation ($\mathcal{L}_{\text{B}}=0$).
Other parameters are the same as those in Fig.~\ref{fig:decay}.}
\label{fig:evolution}
\end{figure*}

In the adiabatic regime ($v/\epsilon^{2}\lesssim1$), $\Gamma_{+}(t)$
is almost a symmetric function of $t$ as shown in Fig.~\ref{fig:decay}(a).
This is because when $\tau_{\text{LZ}}\gg\tau_{B}$, we can approximately
rewrite Eq.~(\ref{eq:G+-}) as
\begin{align}
\Gamma_{+}(t) & \approx\cos^{2}\theta_{t}J\left[2E_{e}(t)\right]\left\{n\left[2E_{e}(t)\right]+1\right\},\label{eq:G+ad}\\
\Gamma_{-}(t) & \approx\cos^{2}\theta_{t}J\left[2E_{e}(t)\right]n\left[2E_{e}(t)\right],\label{eq:G-ad}
\end{align}
where $n(\omega)=1/[\exp(\beta\omega)-1]$ and the expressions on the
right are obvious even functions of time. Equations~(\ref{eq:G+ad},\ref{eq:G-ad})
are usually directly employed in the adiabatic master equation \citep{2014.liu,2018.Xu},
but they are no longer justified if $v$ rises into
the nonadiabatic regime and should be derived on the basis of Eq.~(\ref{eq:G+-}).
As shown in Fig.~\ref{fig:decay}(c) with $v/\epsilon^{2}=10$, two
peaks of $\Gamma_{+}(t)$ becomes asymmetric and $\Gamma_{+}(t)$ slightly
oscillates on top of an exponential decay when $t\apprge2\tau_{\text{LZ}}$
at $T=25\epsilon$. In this nonadiabatic case, $\Gamma_{+}(t)$
becomes negative for a short time as soon as $t>0$. The appearance
of negative $\Gamma_{+}(t)$ means the relaxation turns into pumping.
However, this anomaly does not have a notable effect as
it only lasts for an instant, and the LZ transition is overwhelming simultaneously.

The dephasing rate $\Gamma_{z}(t)$ is approximately proportional
to $\sin^{2}\theta_{t}$ and contributes along with $\vert\alpha_{eg}(t)\vert$
in the time domain. The effect of this term is to reduce the coherence
between the states $\vert e(t_{0})\rangle$ and $\vert g(t_{0})\rangle$,
or equivalently that between the diabatic basis $\vert\uparrow\rangle$ and
$\vert\downarrow\rangle$ as we set the initial evolution time $t_{0}\ll0$
in the following calculation. The energy-level shift $S_{\pm}(t)$
is not shown here, considering its effect on the dynamics is
negligible if the system--bath coupling is weak. 

The time evolution of the populations $P_{\zeta}(t)=\langle\zeta\vert\rho_{S}(t)\vert\zeta\rangle\ (\zeta=\uparrow,\downarrow)$
are shown in Fig.~\ref{fig:evolution}. The sweep velocities of the
subplots in each row are $v/\epsilon^{2}=0.1,1,$ and $10$ from top
to bottom. The left (right) panel shows the high (low)
temperature cases with $T/\epsilon=25$ ($T/\epsilon=1$). The time
evolution starts at $t_{0}=-40\tau_{\text{LZ}}$ with the initial
state $\vert\psi(t_{0})\rangle=\vert\uparrow\rangle$ and ends
at $t_{f}=40\tau_{\text{LZ}}$. Outside this period, the dynamics becomes
trivial because everything decays to zero as shown in Fig.~\ref{fig:decay}.
We also plot two sets of curves for comparison; i.e., $P_{\uparrow(\downarrow),\mathcal{L}_{\text{0}}=0}$
represents the population calculated by manually closing the nonadiabatic
transitions $\mathcal{L}_{\text{0}}[\tilde{\rho}_{S}(t)]=0$ while
$P_{\uparrow(\downarrow),\mathcal{L}_{\text{B}}=0}$ represents the population
in the original LZ problem without dissipation ($\mathcal{L}_{\text{B}}[\tilde{\rho}_{S}(t)]=0$).

As shown in Fig.~\ref{fig:evolution}(a)(d) with $v/\epsilon^{2}=0.1$
in the adiabatic regime, $P_{\uparrow(\downarrow)}(t)$ coincides
well with the curve $P_{\uparrow(\downarrow),\mathcal{L}_{\text{0}}=0}(t)$,
because the dissipation $\mathcal{L}_{\text{B}}[\tilde{\rho}_{S}(t)]$
dominates the time evolution and the LZ transition can be neglected.
When the bath temperature is high [as $T/\epsilon=25$ in Fig.~\ref{fig:evolution}(a)],
the thermal excitation prevents $\vert\uparrow\rangle$ from completely
transferring into $\vert\downarrow\rangle$, and $P_{\downarrow}(t_{f})$
is thus smaller than $1$, which is different from the original unitary LZ
problem. However, if the temperature is low as $T/\epsilon=1$,
the heat bath cools the system to a new ground state after
passing the avoided crossing point, whose effect coincides with
the adiabatic evolution. Thus, all the three sets of curves overlap
in Fig.~\ref{fig:evolution}(d).

The effects of nonadiabatic LZ transition emerge when the
sweep velocity gradually increases to $v/\epsilon^{2}=1$. As shown in Fig.~\ref{fig:evolution}(b)(e),
$P_{\uparrow(\downarrow)}(t>0)$ is slightly higher (lower) than $P_{\uparrow(\downarrow),\mathcal{L}_{\text{0}}=0}(t>0)$
because the LZ transition prefers to reserve more population in
the initial state $\vert\uparrow\rangle$. A fast oscillation with
small amplitude appears on $P_{\uparrow(\downarrow)}(t>0)$; this
is the typical signature of the LZ dynamics. If we further increase
the sweep velocity to $v/\epsilon^{2}=10$, the dynamics of the dissipative
LZ model are similar to those of the original LZ model because of the
relatively weak dissipation. The small disparity between $P_{\uparrow(\downarrow)}(t>0)$
and $P_{\uparrow(\downarrow),\mathcal{L}_{\text{B}}=0}(t>0)$ in Fig.~\ref{fig:evolution}(d) when
$T/\epsilon=25$ is evidence of thermal excitation, which is barely noticeable in the low-temperature case [Fig.~\ref{fig:evolution}(f)].

We emphasize that the dissipative LZ system generally cannot
finally relax to the instantaneous thermal equilibrium state $\rho_{S}^{\text{eq}}(t)\equiv e^{-H_{S}(t)/T}/\text{Tr}[e^{-H_{S}(t)/T}]$, because the dissipation vanishes when the time-dependent energy gap is
much larger than the bath cut-off frequency $\omega_{c}$. As a result,
the system will evolve into a steady state with an effective temperature
higher than that of the heat bath. If we ignore the coherence, we can define
a time-dependent effective temperature of the system as $T_{\text{eff}}(t)=2E_{e}(t)/\ln[P_{\downarrow}(t)/P_{\uparrow}(t)]$,
which increases with the growing $v$. $T_{\text{eff}}(t_{f})$ even becomes negative when $v/\epsilon^{2}=10$,
implying the populations are inverted; i.e., $P_{\downarrow}(t_{f})<P_{\uparrow}(t_{f})$.

\subsection{LZ probability}

\begin{figure}[h]
\includegraphics[width=8cm]{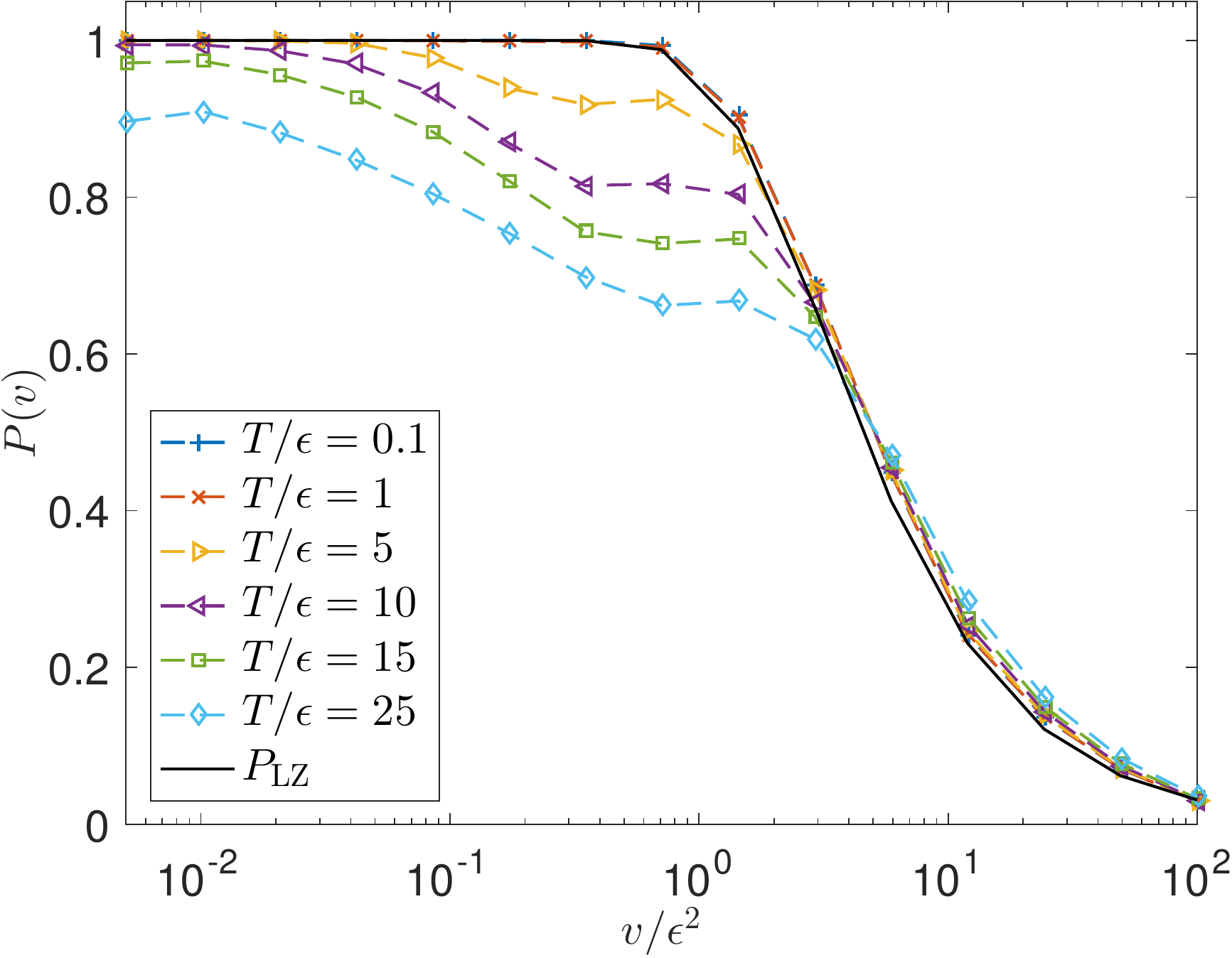}\caption{LZ probability $P(v)$ for different temperatures. The TDQME results
are presented by color markers for $T/\epsilon=0.1$, $1$, $5$,
$10$, $15$, and $25$ while the black solid line is the LZ probability
$P_{\text{LZ}}(v)$. $P(v)$ and $P_{\text{LZ}}(v)$
coincide well when $v/\epsilon^{2}<1$ and the temperature is low
($T/\epsilon=0.1$ and $1$). When the sweep velocity is high ($v/\epsilon^{2}\apprge10$),
the nonadiabatic transition dominates the evolution; hence, $P(v)$
at different temperatures approaches but is a little higher
than $P_{\text{LZ}}(v)$. Such small disparities are the results of dissipation and finally vanish at extremely high
$v$. The nonmonotonic behavior of $P(v)$ appears around $v/\epsilon^{2}=1$
and $T/\epsilon\apprge1$, when the nonadiabatic transition and dissipation
are comparable in strength. Other parameters are the same as those in Fig.~\ref{fig:decay}.}
\label{fig:LZP}
\end{figure}

The LZ probability $P_{\text{LZ}}(v)$ lies at the heart of the LZ
problem and is defined by 
\begin{equation}
P_{\text{LZ}}(v)=\vert\langle\downarrow\vert\mbox{T}_{\leftarrow}e^{-i\int_{-\infty}^{\infty}dtH_{S}(t)}\vert\uparrow\rangle\vert^{2}.\label{eq:PLZ}
\end{equation}
If we initially prepare the two-level system in the state $\vert\uparrow\rangle$
at $t_{0}\rightarrow-\infty$ and then let the system unitarily evolve
to $t_{f}\rightarrow\infty$ governed by linear driving $H_{S}(t)$,
$P_{\text{LZ}}(v)$ is the probability of finally finding the system
in state $\vert\downarrow\rangle$; $P_{\text{LZ}}(v)$ has the well-known expression
\citep{1932.Zener,1932.Landau} as
\begin{equation}
P_{\text{LZ}}(v)=1-\exp\left(-\frac{\pi\epsilon^{2}}{v}\right).\label{eq:PLZ2}
\end{equation}

It is not only interesting but also of practical importance to study
how the dissipation interferes with the LZ transition; indeed, the open-system
version of LZ probability has received broad and long-lasting interest
\citep{1988.Wilkinson,1991.Rammer,1997.Nakayama,2007.Hanggi,2009.Thorwart,2014.Segal,2015.Thorwart,2020.Chen}.
In the case of the open system, we can analogously define the LZ probability $P(v)$
by replacing the unitary evolution in Eq.~(\ref{eq:PLZ}) by the
dissipative dynamics described by the TDQME with initial state $\rho_{S}(t_{0})=\vert\uparrow\rangle\langle\uparrow\vert$; i.e.,
\begin{equation}
P(v)=\langle\downarrow\vert\rho_{S}(t_{f})\vert\downarrow\rangle.\label{eq:Pv}
\end{equation}

We calculate $P(v)$ with $t_{f}=-t_{0}\approx100\tau_{\text{LZ}}$,
such that the time evolution is long enough to cover all the nonadiabatic
and dissipative processes. In Fig.~\ref{fig:LZP}, we plot both $P(v)$
at various temperatures and $P_{\text{LZ}}(v)$ with respect to the logarithmic value of $v$. Except for the low-temperature
cases of $T/\epsilon<1$, in the adiabatic regime, $P(v)$ decreases
with increasing $v$. As we mentioned in the last subsection, the
relaxation is active only on a time scale inverse to $v$; thus, the system
decouples with the heat bath at an earlier time as $v$ increases before it is in equilibrium,
and greater populations remain in $\vert\uparrow\rangle$. As the sweep velocity further increases entering
the region $v/\epsilon^{2}\sim1$, the gradually enhanced LZ transition
starts to compete with dissipation and eventually dominates the evolution.
Therefore, a local minimum of $P(v)$ appears around $v/\epsilon^{2}\sim1$
and then $P(v)$ quickly decreases with $v$ increasing into the LZ regime
($v/\epsilon^{2}\gg1$). The residual effect of the heat bath in the regime $1\lesssim v/\epsilon^{2}\lesssim10$ is reflected by the small disparities of $P(v)$ above $P_{\text{LZ}}(v)$.

We note that the nonmonotonicity of the LZ probability
is more apparent in the case of longitudinal system--bath coupling than
in the case of transverse coupling discussed here, where the
relaxation plays a more important role (see Appendix~\ref{sec:A}). Our results obtained
using the TDQME method are well consistent with those calculated using the QUAPI \citep{2015.Thorwart}. 

\section{Finite-time thermodynamics}

The above calculation suggests that the TDQME is a useful tool with which to explore
the finite-time thermodynamics of quantum systems with discrete energy
levels. In the framework of quantum thermodynamics, the internal energy $U(t)=\textrm{Tr}\left[\rho_{S}(t)H_{S}(t)\right]$
and von Neumann entropy $S(t)=-\text{Tr}[\rho_{S}(t)\ln\rho_{S}(t)]$
are functions of the system state at the present time. However, the input work $W(t)$ and the absorbed heat $Q(t)$ depend on the driving protocol. The work is defined as
\begin{equation}
W(t)=\int_{t_{0}}^{t}d\tau\textrm{Tr}\left[\rho_{S}(\tau)\frac{\partial H_{S}(\tau)}{\partial\tau}\right],\label{eq:W}
\end{equation}
and the heat can be obtained using the first law of thermodynamics as $Q(t)=\Delta U(t)-W(t)$,
where $\Delta U(t)=U(t)-U(t_{0})$ is the internal energy increase. 

The second law of thermodynamics tells us that the entropy does not decrease
in a closed system. Accordingly, the total
entropy production, usually termed as irreversible entropy production
$\Delta S_{\text{irr}}(t)$, is always nonnegative and can be decomposed as
\begin{equation}
\Delta S_{\text{irr}}(t)=\Delta S(t)-\Delta S_{\text{e}}(t)\geq0.\label{eq:Sirr}
\end{equation}
Here, $\Delta S(t)=S(t)-S(t_{0})$ and $\Delta S_{\text{e}}(t)\equiv Q(t)/T$
is the entropy decrease of the heat bath due to the output heat current.
In finite-time processes, the system is generally in
a nonequilibrium state, and the irreversible entropy production
$\Delta S_{\text{irr}}(t)$ serves to characterize the deviation of
the system away from equilibrium \citep{2009.Jarzynski,2011.Broeck}.

\begin{figure*}
\includegraphics[width=13cm]{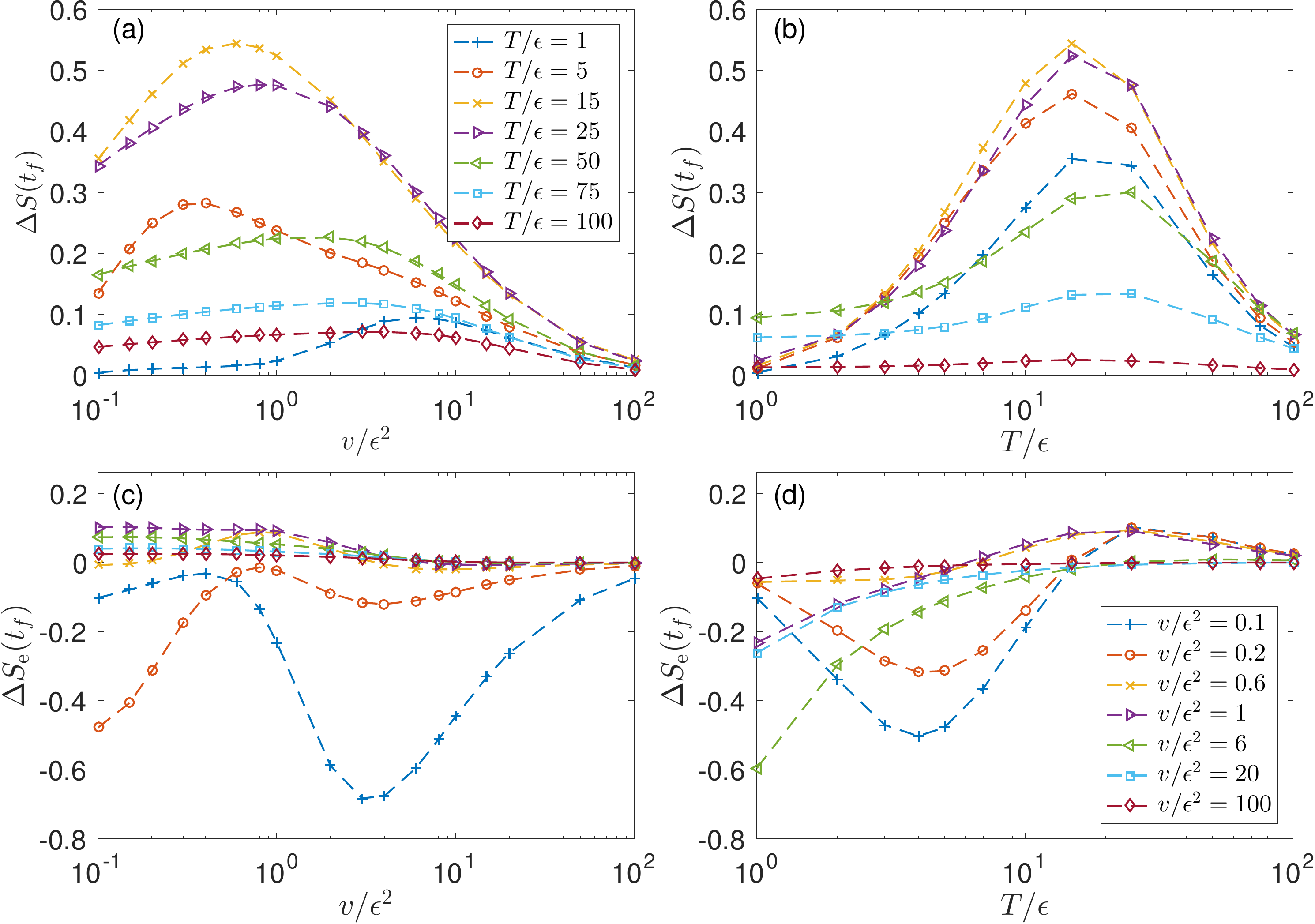}\caption{Changes in entropy $\Delta S(t_{f})$ and entropy flow $\Delta S_{\text{e}}(t_{f})$ as functions of $v$ in subplots (a)(c) and $T$ in subplots (b)(d).
The initial state is the instantaneous equilibrium state at $t_{0}=-40\tau_{\text{LZ}}$. Other parameters are the same as those in Fig.~\ref{fig:decay}.}
\label{fig:QandS}
\end{figure*}

It is usually believed that the leading order of $\Delta S_{\text{irr}}(t)$
is inversely proportional to the total evolution time under the low-dissipation assumption ( 
i.e., $\Delta S_{\text{irr}}(t_{f})\propto1/(t_{f}-t_{0})$
\citep{2010.Broeck}), which has been confirmed both theoretically \citep{2017.Giovannetti,2018.Xu}
and experimentally \citep{2020.Dong} in systems without energy
level avoided crossing. The validity of the low-dissipation assumption
requires there should be no retard of equilibrium, which readily occurs in fast-driving processes. To explore the behavior
of $\Delta S_{\text{irr}}(t)$ in a broader region, we start with
the calculation of $\Delta S(t)$ and $\Delta S_{\text{e}}(t)$ in
the dissipative LZ model. We initialize the system in a thermal equilibrium
state $\rho_{S}(t_{0})=e^{-\beta H_{S}(t_{0})}/\text{Tr}[e^{-\beta H_{S}(t_{0})}]$,
and then let the system evolve from $t_{0}=-40\tau_{\text{LZ}}$ to
$t_{f}=40\tau_{\text{LZ}}$, when the system finally approaches a
nonequilibrium steady state.

In the high-sweep-velocity region ($v/\epsilon^{2}\apprge10$), the
LZ system evolves in an almost unitary way under the slight effect of 
the heat bath. As a result, $\Delta S(t_{f})$ and $\Delta S_{\text{e}}(t_{f})$
quickly approach zero with increasing $v$, as shown in Fig.~\ref{fig:QandS}(a)
and (c). Even when the temperature is much higher than $\epsilon$ (e.g., $T/\epsilon=100$), the LZ transition is still hundreds
of times larger in magnitude than the dissipation during $t\in[-\tau_{\text{LZ}},\tau_{\text{LZ}}]$;
hence, the dissipation can be safely neglected. Meanwhile,
when the sweep velocity is low (e.g., $v/\epsilon^{2}\apprle1$), the
LZ transition becomes negligible, such that the system tends to
adiabatically follow its instantaneous eigenstates while exchanging
heat with the bath. A lower sweep velocity means that more population
remains in the initial adiabatic basis, and the change in system entropy
$\Delta S(t_{f})$ decreases accordingly. In the intermediate region
$1\apprle v/\epsilon^{2}\apprle10$, the LZ transition and the adiabatic
following compete with each other, which largely disturbs the population
distribution and results in local maximums of $\Delta S(t_{f})$.

Local maximums of $\Delta S(t_{f})$ also appear when we change temperature
and fix the sweep velocity [Fig.~\ref{fig:QandS}(b)]. In the
low-temperature limit as $T/\epsilon$ approaching $1$, the dissipation
is too weak to substantially affect the unitary evolution,
and $\Delta S(t_{f})$ thus decreases to a small value. At high-temperature
limit as $T/\epsilon\rightarrow100$, the critical issue of small $\Delta S(t_{f})$ rises from the fact that the entropy of a
binary distribution changes less when the entropy is greater. Consequently, no matter how much the initial state populations
are changed by the subsequent dynamics, $\Delta S(t_{f})$ is bounded
by $\ln2-S(t_{0})$, which is small for a high-temperature initial equilibrium state. 

\begin{figure*}
\includegraphics[width=13cm]{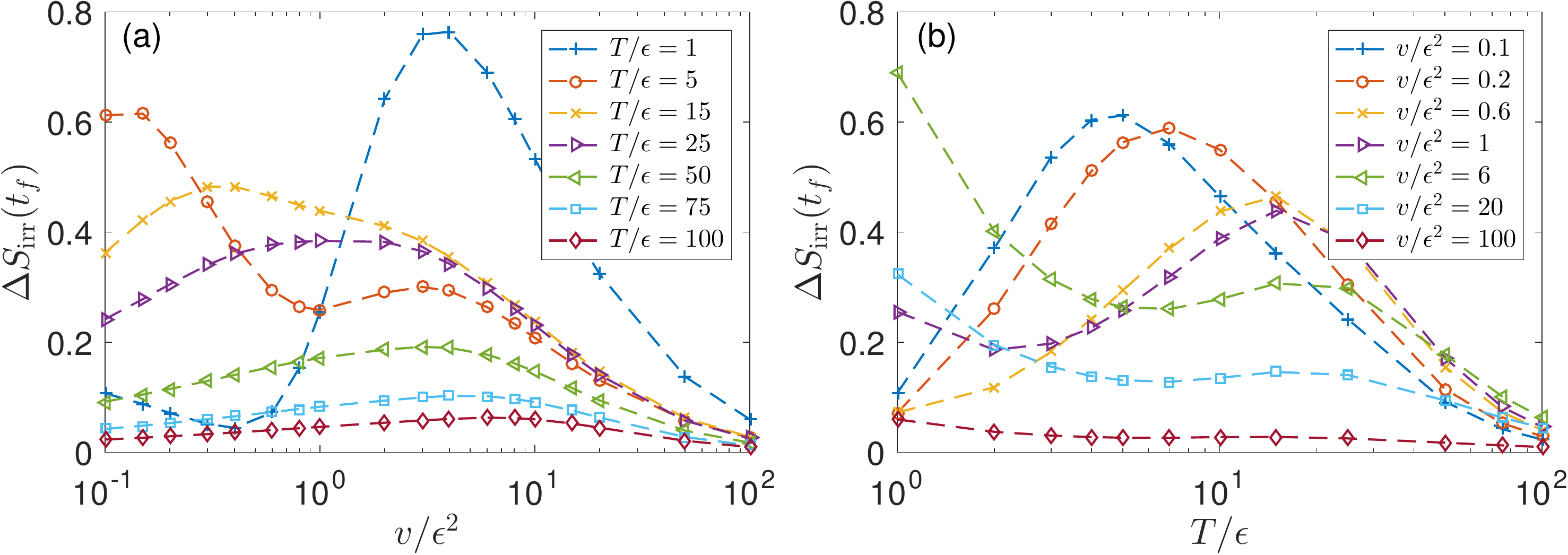}\caption{Irreversible entropy production $\Delta S_{\text{irr}}(t_{f})$
as functions of (a) $v$ and (b) $T$. The low-dissipation regime
has low sweep velocities and high temperatures, where $\Delta S_{\text{irr}}(t_{f})$
changes linearly with $v$. Other parameters are the same as those in Fig.~\ref{fig:QandS}.}
\label{fig:Sirr}
\end{figure*}

The entropy flow $\Delta S_{\text{e}}(t_{f})$ as functions of the sweep
velocity and temperature is presented in Fig.~\ref{fig:QandS}(c)
and (d), respectively. At the high sweep velocity limit or high temperature
limit, $\Delta S_{\text{e}}(t_{f})$ is extremely small for a reason similar to that for $\Delta S(t_{f})$. We note that $\Delta S_{\text{e}}(t_{f})$ is small at the high $T$ limit also
because $\Delta S_{\text{e}}(t_{f})$ equals the heat absorption rescaled
by $1/T$. Furthermore, the heat absorption is positively relates to the variation
rate of the population difference between two instantaneous eigenstates,
which is greatly suppressed at high $T$. More sophisticated behavior
of $\Delta S_{\text{e}}(t_{f})$ appears at low temperature ($T/\epsilon\apprle10$)
in the intermediate sweep velocity region, which is consistent with the nonmonotonic region of $P(v)$ in Fig.~\ref{fig:LZP},
where the LZ transition is maximally affected by dissipation. 

The irreversible entropy production $\Delta S_{\text{irr}}(t_{f})$
is plotted as functions of $v$ and $T$ in Fig.~\ref{fig:Sirr}(a)
and (b), respectively. Its behavior can be understood by combining
the above results, which are not repeated here. As shown in Fig.~\ref{fig:Sirr}(a),
$\Delta S_{\text{irr}}(t_{f})$ linearly increases with $v$ only in the
region of low $v$ ($v/\epsilon^{2}<1$) and high $T$ ($T/\epsilon>50$), beyond which $\Delta S_{\text{irr}}(t_{f})$ has a  nonmonotonic dependence on $v$. 
Being linearly proportional to $v$ is equivalent to being inversely proportional to $t_f-t_0$ in the current situation, 
which indicates the low-dissipation regime. An important
aspect of the TDQME is that regardless of the low-dissipation assumption, we can now quantitatively determine under what conditions $\Delta S_{\text{irr}}(t_{f})$ reaches a maximum or minimum, which lies at the heart of designing highly
efficient thermodynamic cycles. It is worth noticing that the energy-level avoided crossing can induce more profound dynamic and thermal phenomena, which should be considered for optimization problems in finite-time thermodynamics.

\section{CONCLUSIONS AND OUTLOOK}

We formally derived a Born--Markovian quantum master equation
for the time-dependent open system in the adiabatic basis representation
based on the Nakajima--Zwanzig projective operator method. Our formalism can be
applied to an arbitrary driving protocol with a broad range of driving
speeds and is justified even beyond the adiabatic regime. Several authors have
pointed out that the nonadiabatic transition and
dissipation hardly dominate the system dynamics simultaneously; thus,
the adiabatic condition is not a necessary requirement in our approach. This idea is explicitly implemented
in our formalism by separating the Liouville operator into the nonadiabatic
transition and the dissipation terms. Taking the dissipative LZ
system as an example, we show that when the system is driven quickly, the nonadiabatic
transition overwhelms the dissipation. When the
sweep velocity is decreasing, the nonadiabatic transition decreases in magnitude;
meanwhile, it separates from the dissipation in the time domain. The
validity of our approach was checked by calculating the LZ probabilities
with finite-temperature heat bathes and comparing with previously obtained numerical
results. 

The competition between the time-dependent dissipation and nonadiabatic
transition plays an essential role in determining the thermodynamic
properties of finite-time processes, which is clearly revealed by
the dissipative LZ system. Our calculation shows that in
the case of a system linearly passing through the avoided crossing point,
the entropy, heat, and irreversible entropy change nonmonotonically
with the sweep velocity and bath temperature. The low-dissipation regime
only appears at high temperature and for slow driving, which is consistent
with the usual understanding. However, our TDMQE is applicable in situations with parameters
beyond the low-dissipation regime and can be used for the problems such as the
optimization of the performance of the quantum Carnot cycle \citep{2018.Sun}, and the distributions
of the work and heat in stochastic quantum thermodynamics \citep{2013.Pekola,2014.liu,2014.Ala-Nissila,2014.Sassetti,2018.Quan,2015.Mottonen}.

\begin{acknowledgments}
D. X. thanks Hui Dong and Jinfu Chen for helpful discussion. This
study is supported by the NSF of China (Grants No. 11705008 and Grants
No. 12075025). 
\end{acknowledgments}

\appendix
\section{LZ probability with longitudinal system-bath coupling~\label{sec:A} }

This appendix presents the results of the dissipative LZ model with
pure longitudinal system--bath coupling
\begin{align}
V & =\sigma_{z}\otimes\sum_{k}g_{k}(a_{k}^{\dagger}+a_{k}),\label{eq:Vz}
\end{align}
which after transformation into the adiabatic frame of reference reads
$\tilde{V}(t)=\tilde{\sigma}_{z}(t)\otimes B(t)$ with
\begin{equation}
\tilde{\sigma}_{z}(t)=\cos\theta_{t}\hat{\sigma}_{z}-\sin\theta_{t}[e^{-i\phi_{eg}(t)}\hat{\sigma}_{+}+\text{h.c.}].\label{eq:Szad}
\end{equation}
Equation~(\ref{eq:Szad}) is the same as Equation~(\ref{eq:Sxad}) for $\tilde{\sigma}_{x}(t)$ except that $\cos\theta_{t}$ and
$\sin\theta_{t}$ are interchanged. As a result, the relaxation only occurs
within the time window $t\leq\vert\tau_{\text{LZ}}\vert$ in the longitudinal
coupling case, while pure dephasing occurs outside the time window.
Comparing with the dissipative rates in Fig.~\ref{fig:decay}, the relaxation
in the transverse coupling case effectively exists for longer time
than that in the longitudinal coupling case, resulting in more apparent
dissipative effects. 

\begin{figure}[pbth]
\includegraphics[width=8cm]{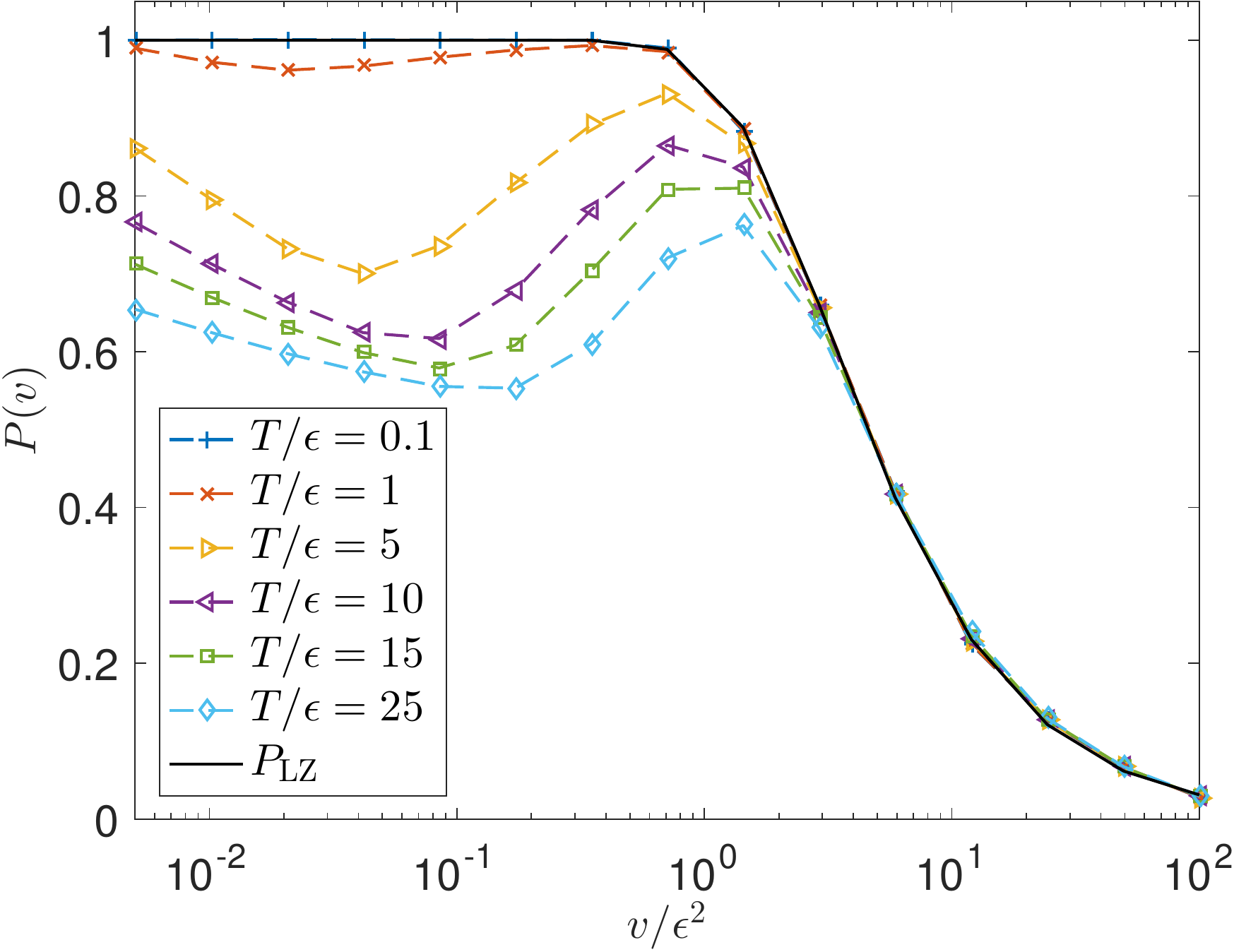}
\caption{LZ probability $P(v)$ of the longitudinal system--bath coupling
case for various temperatures. The TDQME results are
presented with color markers for $T/\epsilon=0.1$, $1$, $5$, $10$,
$15$, and $25$, while the black solid line is the LZ probability
$P_{\text{LZ}}(v)$. $P(v)$ and $P_{\text{LZ}}(v)$
coincide well when $v/\epsilon^{2}\gtrsim2$ for all the temperatures.
Other parameters are the same as those in Fig.\ref{fig:LZP}.}
\label{fig:Pz}
\end{figure}

We plot the LZ probability of the longitudinal system--bath coupling case in Fig.~\ref{fig:Pz}. There are two main differences from the transverse coupling
case: the local minimums of $P(v)$
have smaller values and are at lower $v$ in the low-sweep-velocity regime and
$P(v)$ converges more quickly to $P_{\text{LZ}}(v)$ when $v/\epsilon^{2}\gtrsim1$.
The simple explanation is that in the longitudinal coupling case,
there is relaxation when the energy gap is small at about $2\epsilon$.
Thus, in the regime of low sweep velocity, the bath intends to relax
the two-level system with more evenly distributed populations on the
instantaneous eigenstates at finite temperature, and $P(v)$
is therefore smaller than that in the transverse coupling case. In the high-sweep-velocity regime, as the nonadiabatic transition and relaxation occur in the same time window and the former overwhelms the later, the effect of the heat bath is totally washed out regardless of temperature. Our results are consistent with the QUAPI
results in Ref.~\citep{2009.Thorwart,2015.Thorwart}.

\bibliographystyle{apsrev4-1}
\bibliography{TDQME}

\end{document}